\documentclass[%
 aip,
 amsmath,amssymb,
 reprint,%
]{revtex4-1}

\usepackage{graphicx}
\usepackage{dcolumn}
\usepackage{bm}

\usepackage[utf8]{inputenc}
\usepackage[T1]{fontenc}
\usepackage{mathptmx}
\usepackage{etoolbox}

\makeatletter
\def\@email#1#2{%
 \endgroup
 \patchcmd{\titleblock@produce}
  {\frontmatter@RRAPformat}
  {\frontmatter@RRAPformat{\produce@RRAP{*#1\href{mailto:#2}{#2}}}\frontmatter@RRAPformat}
  {}{}
}%
\makeatother
\begin{document}

\preprint{AIP/123-QED}

\title{High harmonic generation in monolayer MoS\textsubscript{2} controlled by resonant and near-resonant pulses on ultrashort time scales}

\author{Pavel Peterka}
\affiliation{Department of Chemical Physics and Optics, Faculty of Mathematics and Physics, Charles University, Ke Karlovu 3, 12116 Prague 2, Czech Republic}
\author{Artur O. Slobodeniuk}
\author{Tomáš Novotný}
\affiliation{Department of Condensed Matter Physics, Faculty of Mathematics and Physics, Charles University, Ke Karlovu 3, 12116 Prague 2, Czech Republic}
\author{Pawan Suthar}
\affiliation{Department of Chemical Physics and Optics, Faculty of Mathematics and Physics, Charles University, Ke Karlovu 3, 12116 Prague 2, Czech Republic}
\author{Miroslav Bartoš}
\affiliation{Central European Institute of Technology, Brno University of Technology, Purkyňova 656/123, 612 00 Brno, Czech Republic}
\author{František Trojánek}
\author{Petr Malý}
\author{Martin Kozák}
\email{kozak@karlov.mff.cuni.cz}
\affiliation{Department of Chemical Physics and Optics, Faculty of Mathematics and Physics, Charles University, Ke Karlovu 3, 12116 Prague 2, Czech Republic}

\date{\today}

\begin{abstract}
We report on experimental investigation of nonperturbative high harmonic generation (HHG) in monolayer MoS\textsubscript{2} in the ultraviolet spectral region driven by mid-infrared light. We study how the HHG is influenced by pre-excitation of the monolayer using resonant and near-resonant pulses in a pump-probe-like scheme. The resonant light creates high density exciton population. Due to ultrafast dephasing caused by electron-electron scattering, the HHG is suppressed in the presence of pre-excited carriers. In the case of near-resonant excitation with photon energy below the exciton transition, the dynamics of the observed suppression of the HHG yield contains a fast component which is a consequence of momentum scattering at carriers, which are excited by two-photon transition when the two pulses temporally overlap in the sample. This interpretation is supported by comparison of the experimental data with theoretical calculations of two-photon absorption spectrum of MoS\textsubscript{2} monolayer. This work demonstrates a possibility to control HHG in low-dimensional materials on ultrashort timescales by combining the driving strong-field pulse with a weak near-resonant light.
\end{abstract}

\maketitle

\section{Introduction}
High harmonic generation (HHG) in solids \cite{ghimire2011observation,luu2015extreme,vampa2015linking,ghimire2019high,yoshikawa2019interband,yoshikawa2017high} is a nonperturbative nonlinear optical process, during which coherent light wave with high amplitude of electric field and low photon energy interacts with a solid-state material, typically a crystal. As a result, electron-hole wave packets are promoted to conduction and valence bands via quantum tunnelling. This process occurs during a short time window close to the maxima of the oscillating field. The duration of the tunnelling excitation window is typically few hundreds of attoseconds to single femtoseconds, depending on the frequency of the driving light. Subsequently, the carriers are coherently accelerated in the crystal by the oscillating field and propagate far from equilibrium positions. The electron and hole can eventually recombine leading to generation of high energy photons via interband mechanism of HHG\cite{vampa2014theoretical,vampa2015linking,ghimire2019high,yoshikawa2019interband}. Due to the nonparabolicity of electron and hole dispersion in the material, the harmonically driven carriers undergo anharmonic motion leading to so-called intraband mechanism of HHG\cite{golde2008high,wu2015high,schubert2014sub}. HHG in solids has been used to investigate band structure \cite{vampa2015all,suthar2022}, Berry curvature\cite{luu2018measurement}, topological surface states\cite{bai2021high}, the dynamics of a photoinduced phase transition\cite{bionta2021tracking} or coherent phonon dynamics\cite{Neufeld2022}.

A crucial requirement for observation of a macroscopic wave containing harmonic frequencies is the coherence of the individual microscopic emission sources which contribute to the radiated field\cite{wang2021quantum,floss2019incorporating}. However, when the charge carriers in a solid are accelerated to high energies, the momentum scattering causes fast decoherence. Most important scattering processes are electron-phonon scattering, electron-electron scattering and scattering at ionized impurities. While the electron-phonon scattering times for electrons with low kinetic energy are typically long (hundreds of femtoseconds to picoseconds) compared to the time period of the mid-infrared light (several femtoseconds), the electron-phonon and electron-electron scattering times can become much shorter for electrons that are accelerated to high kinetic energies and/or at high densities of excited carriers. The latter process can be controlled independently of the HHG process using a second resonant pulse exciting the carriers prior to the impact of the strong infrared pulse, which drives the HHG. The suppression of the HHG yield by excited carriers was recently observed in ZnO\cite{wang2017roles}, carbon nanotubes\cite{Nishidome2020} or MoS\textsubscript{2} monolayer\cite{heide2022probing,nagai2021effect,wang2022optical}, where the authors used infrared pulses to drive HHG. The carriers were excited by a resonant pump pulse \cite{wang2017roles,nagai2021effect,heide2022probing} or by a pulse with photon energy higher than the bandgap\cite{wang2022optical}. HHG in ZnO was also modulated by pulses with photon energy lower than the band gap\cite{xu2022ultrafast} and the ultrafast response was interpreted as a consequence of nonlinear frequency mixing present in the sample for zero time delay between the two pulses.

In this paper we show an experimental study of the effect of resonant and near-resonant pre-excitation on the HHG process in monolayer MoS\textsubscript{2}\cite{mak2010atomically}. The harmonics are generated in a different regime than in previous studies\cite{yoshikawa2019interband,liu2017high,heide2022probing,nagai2021effect,wang2022optical,cao2021inter,kobayashi2021polarization}. We use higher photon energy of the driving pulse of 0.62 eV (wavelength of 2000 nm) and observe harmonic spectra in the ultraviolet spectral region 2.7-6.2 eV (wavelength region of 200-450 nm). These photon energies correspond to interband transitions between higher lying bands in the MoS\textsubscript{2} band structure\cite{yue2022signatures}, thus going beyond the two-band approximation. We observe that the HHG yield is suppressed by high density carriers excited in the sample prior or during the illumination by the strong infrared pulse, which drives the HHG. The dynamics of the change of the HHG yield as a function of the time delay between the pre-excitation and driving pulses differs significantly depending on the photon energy of the pre-excitation pulse. For non-resonant pre-excitation, the observed ultrafast supression of the HHG yield is explained by momentum scattering of coherently driven electron-hole wavepackets at charge carriers produced by two-photon absorption induced simultaneously by the pre-excitation and the strong-field pulses. The spectrum of two-photon absorption in MoS\textsubscript{2} is theoretically calculated and compared to the measured results. 

 
\section{Experimental}

\begin{figure*}[ht!]
\centering\includegraphics[width=15cm]{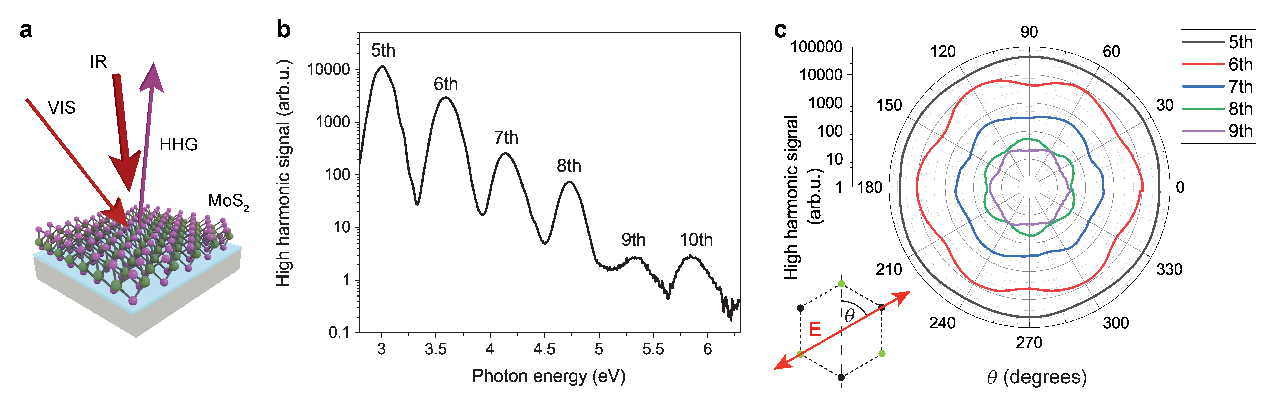}
\caption{(a) Geometry of the experiment for investigation of the role of resonant and near-resonant excitation on the yield of high harmonic generation (HHG) in MoS\textsubscript{2} monolayer. (b) HHG spectrum generated in the MoS\textsubscript{2} monolayer with the linear polarization of the driving field along the mirror symmetry axis of the crystal (armchair direction). (c) Dependence of the polarization unresolved HHG yield of each harmonic frequency on the orientation of the linear polarization of the driving field with respect to the mirror symmetry axis of the crystal.}
\label{fig:Figure1}
\end{figure*}

In our experiments we illuminate a MoS\textsubscript{2} monolayer on SiO\textsubscript{2} substrate with mid-infrared pulses and measure the spectra of high harmonic frequencies reflected from the sample (see Fig. \ref{fig:Figure1}a). The HHG spectra are measured as a function of the time delay between a femtosecond resonant or near-resonant pump pulse (tunable photon energy 1.25-2 eV, pulse duration of about 20 fs) and the mid-infrared high intensity pulse (central photon energy 0.6 eV, FWHM pulse duration of 38 fs, peak intensity up to 5 TW/cm\textsuperscript{2}), which drives the HHG. Both pulses are generated from the output of a femtosecond ytterbium-based laser (Pharos SP-6W, Light Conversion, 170 fs, 1030 nm). The resonant and near-resonant pulses are generated in an in-house developed noncollinear optical parametric amplifier (NOPA), which uses femtosecond supercontinuum generated in a sapphire crystal as a seed, and which is pumped by the second harmonics of the fundamental laser output (515 nm) generated in a BBO crystal. The mid-infrared pulses used to drive the HHG are obtained by the NOPA+DFG (difference frequency generation) setup, which is described in detail in\cite{kozak2021generation}. The time delay between the two pulses is controlled using a motorized translation stage. During the experiments, the samples are imaged in situ using an optical microscope setup to ensure the spatial overlap of the pre-excitation and the driving pulses and their position at the sample. We use linear polarizations of both pulses, which can be independently controlled by broadband half wave plates. Our HHG detection setup has different sensitivities for two orthogonal polarization components due to the prisms used to filter out the third harmonic signal and due to the polarization dependence of the diffraction efficiency of the spectrometer grating. For this reason, in the measurement showing the dependence of the HHG yield on the direction of the polarization of the driving field we detect the harmonic spectra separately for the two orthogonal polarization components, which are selected by a UV Glen-laser polarizer. The signals are corrected for different detection sensitivity and the total HHG yield is obtained as their sum. The experiments are carried out at room temperature with the laser repetition rate of 25 kHz.
The studied samples are prepared by gel-film assisted mechanical exfoliation from bulk MoS\textsubscript{2} crystal. The monolayers are then transferred to a SiO\textsubscript{2} substrate, which is used due to its wide band gap leading to a negligible contribution to the generated harmonic spectra.

\section{Results and discussion}
The measured HHG spectrum for linear polarization of the driving pulse along the direction of the mirror symmetry axis of the MoS\textsubscript{2} monolayer is shown in Fig.~\ref{fig:Figure1}b. We observe harmonic frequencies up to the 10th order corresponding to the photon energy of 6 eV. This is considerably higher photon energy than observed in previous studies of HHG in TMD materials \cite{yoshikawa2019interband,liu2017high,heide2022probing,nagai2021effect,wang2022optical}. The reason is probably the shorter wavelength of the driving field compared to previous experiments in combination with high field strength of about 3 GV/m used in our experiments, which can be applied to the monolayer due to the short pulse duration. The yield of HHG in MoS\textsubscript{2} monolayer and its polarization state is found to depend on the direction of linear polarization with respect to the mirror symmetry plane \cite{liu2017high}. In Fig.~\ref{fig:Figure1}c we plot the HHG yield as a function of the angle $\theta$ between the direction of linear polarization of the driving light and the mirror symmetry plane of the monolayer. The two perpendicular polarization components (horizontal and vertical) of HHG radiation are detected separately. The resulting spectrum is obtained by summing the two polarization components with particular weights corresponding to the measured polarization dependence of the detection setup (prisms, spectrometer grating). The results thus represent the total HHG yield independent of the polarization state of the generated ultraviolet light. We observe harmonic orders 5-9 at photon energies of 2.8-5.5 eV (higher photon energies are not transmitted through the polarizer used in the detection setup for these measurements). For photon energies below 4 eV, the maxima of odd orders are observed for $\theta=30^\circ$, while the even orders have their maxima for $\theta=0^\circ$. However, at higher photon energies, the maximum yield of even and odd orders switches. This behaviour has been previously observed in MoS\textsubscript{2} monolayer and it was attributed to the role of higher energy bands in the HHG process \cite{yue2022signatures}.

\begin{figure*}[ht!]
\centering\includegraphics[width=15cm]{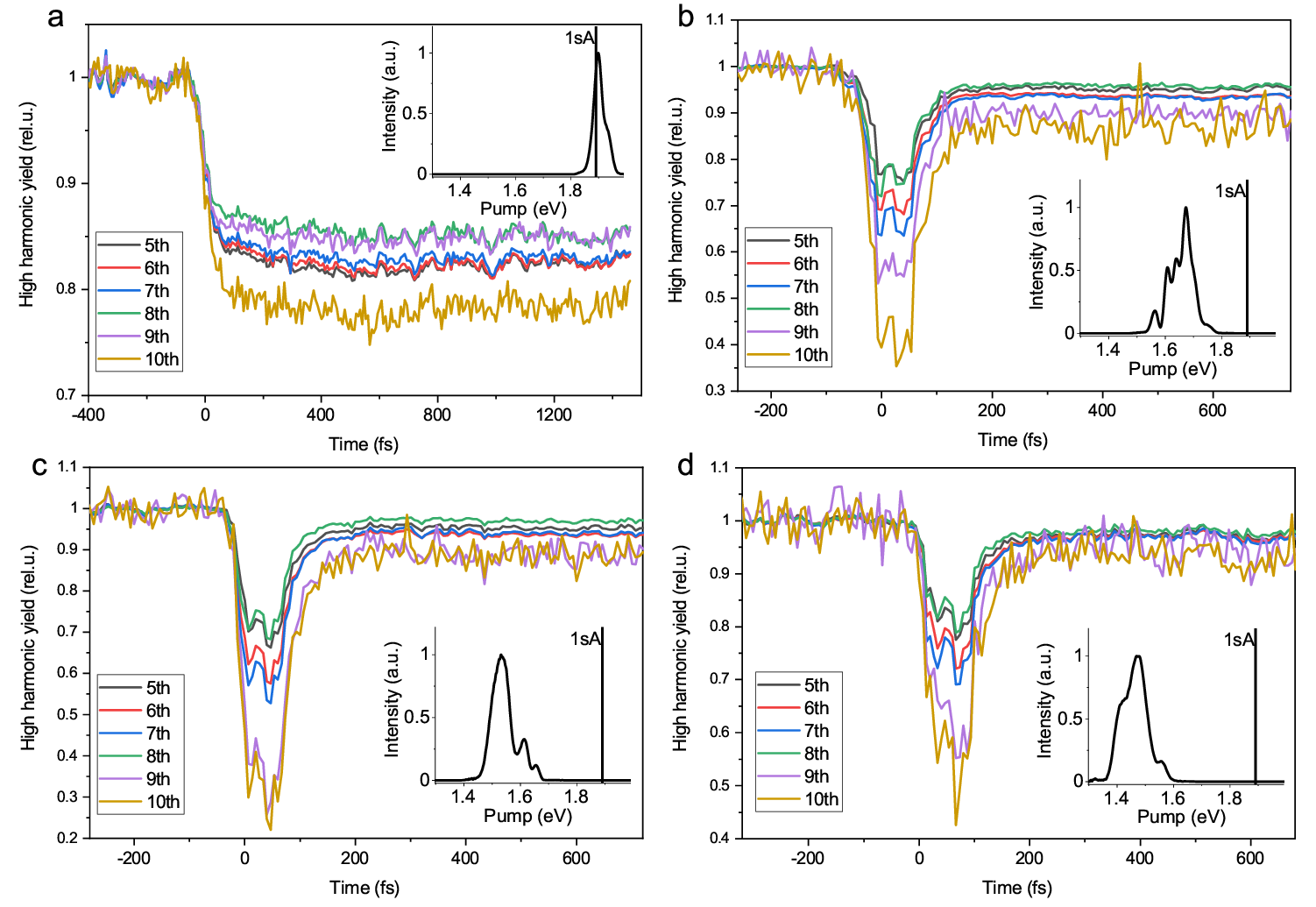}
\caption{(a) Relative change of the high harmonic yield for different harmonic orders as a function of the time delay between the resonant pre-excitation at photon energy of 1.91 eV and the infrared pulse. (b), (c), (d) The same as in (a) with the pre-excitation pulse detuned from the exciton resonance to photon energies of (b) 1.65 eV, (c) 1.55 eV, (d) 1.46 eV. The spectra of the excitation pulses are shown in the insets, where the energy of the lowest exciton resonance (1sA exciton) is indicated.}
\label{fig:hhgtime}
\end{figure*}

The main goal of this paper is to study the influence of the HHG yield by a resonant or near-resonant pump pulse applied to the sample prior to the illumination by the strong infrared pulse which drives the HHG process. When the real carrier population is generated in the sample by a resonant pump pulse, the HHG yield decreases for all the harmonics, which is shown in Fig.~\ref{fig:hhgtime}a. An interesting observation is the fact that the relative decrease is not monotonically changing with the harmonic order as observed in previous studies for harmonic photon energies up to 3.5 eV\cite{heide2022probing,nagai2021effect,wang2022optical}. The decrease of the HHG yield is caused by electron-electron scattering induced by the high density of excited carriers, which leads to extremely fast dephasing of the coherent electron-hole wave packets\cite{heide2022probing}. The generated high harmonic radiation results from the macroscopic nonlinear current of coherently oscillating electrons. Once the quantum phase between the electron and the field of the driving infrared wave is lost, the electron does not further contribute to the generated coherent wave. As the density of the excited carriers decreases on picosecond time scales\cite{heide2022probing,nagai2021effect}, the harmonic signal recovers to the value without the resonant pump after several tens of picoseconds. 

\begin{figure*}[ht!]
\centering\includegraphics[width=14cm]{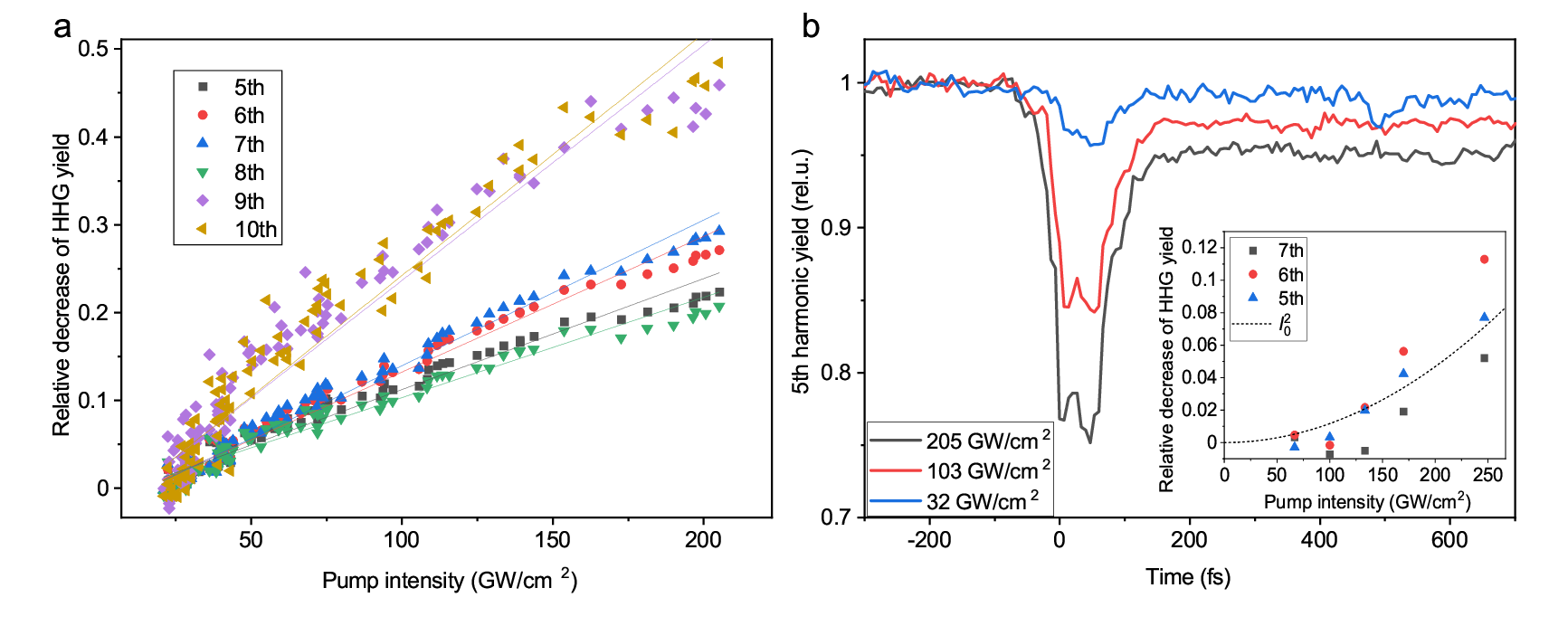}
\caption{(a) Relative decrease of HHG yield for different harmonic orders as a function of the intensity of the off-resonant pump pulse at energy 1.46 eV. Measurements are performed for the two pulses overlapped on the sample (time delay 0 fs). (b) 5th harmonic yield as a function of the time delay between the pre-excitation and the infrared pulses for three different pump pulse intensities. Inset shows relative decrease of HHG yield 500 fs after pre-excitation for different  harmonic orders as a function of the intensity of the off-resonant pump pulse.}
\label{fig:powerdep}
\end{figure*} 

Unlike the other experiments in MoS\textsubscript{2} we also focus on the experiments with the pump beam at photon energy below the lowest exciton resonance of MoS\textsubscript{2} (data shown in Fig.~\ref{fig:hhgtime}b-d). When the near-resonant pulse overlaps in time with the strong infrared pulse, the harmonic yield is significantly decreased. In this case the strong suppression shows ultrafast dynamics and harmonic yield recovers after 80 fs to value that is close to the yield without the pump pulse. Figure~\ref{fig:powerdep}a shows the maximum relative change of the harmonic emission yield as a function of the intensity of the off-resonant pulse for overlapping pump and infrared pulses (time delay 0 fs). The data are fitted by linear function. The harmonic intensity suppression is different for different harmonic orders, it is smallest for 8th harmonic and for 5th, 6th, 7th, 9th and 10th it is monotonically increasing. After the short signal suprression, there is also a picosecond component of the change of HHG yield in the case of near-resonant excitation (see Fig.~\ref{fig:powerdep}b). This signal cannot be caused by the excitons generated by single-photon absorption process due to the fact that the photon energy is lower than the energy of the exciton resonance. The picosecond component is thus caused by carriers excited by two-photon absorption of the near-resonant pulse, which is confirmed by the quadratic dependence of its amplitude shown in the inset of Fig. \ref{fig:powerdep}b.  

\begin{figure*}[ht!]
\centering\includegraphics[width=13cm]{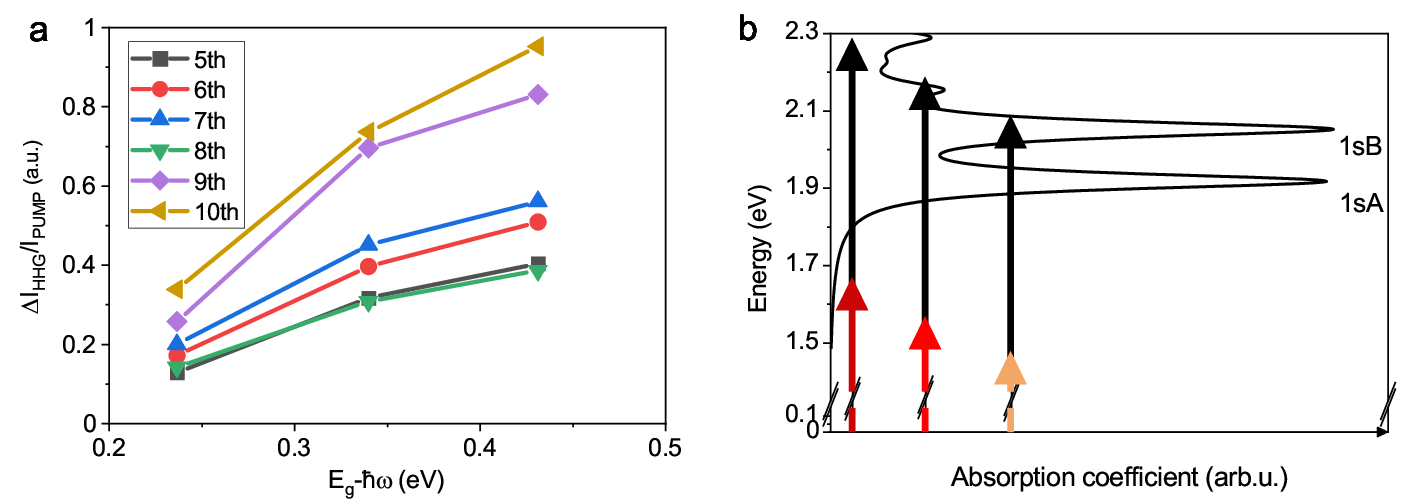}
\caption{(a) The magnitude of the relative decrease of HHG yield as a function of detuning between the photon energy of the near-resonant pulse and the energy of the 1sA exciton resonance. The change of the HHG yield is normalized to the peak intensity of the near-resonant pump pulse. (b) Sketch of two-photon generation of excitons for three different photon energies of the near-resonant pulse of 1.65 eV, 1.55 eV and 1.46 eV.}
\label{fig:absorption}
\end{figure*}

The ultrafast suppression of the HHG yield might be attributed to two effects. The first possible source is the optical Stark effect (OSE)\cite{autler1955stark,ritus1967shift,ell1989influence,chemla1989excitonic,delone1999ac} induced by the near-resonant field. OSE was previously observed in 2D materials\cite{kim2014ultrafast,sie2015valley,sie2017large,lamountain2018valley,cunningham2019resonant,slobodeniuk2022semiconductor,slobodeniuk2022giant} and leads to a transient shift of the excitonic resonance. The energy shift induced by OSE may increase the energy barrier that the electrons need to overcome during tunnel excitation, which is an important ingredient of HHG. As a consequence, the tunnelling probability would be reduced leading to suppression of the harmonic yield. In two-level approximation representing a simplified model of the excitonic system in 2D transition metal dichalcogenides, the OSE induced energy shift scales with the field amplitude $F$, energy of the resonance $E_0$ and the applied photon energy $\hbar\omega$ as $\Delta E \approx\frac{|F|^2}{E_0-\hbar\omega}$. A theory of OSE based on perturbative solution of semiconductor Bloch equations gives only a small correction to this dependence\cite{slobodeniuk2022semiconductor}. The second possible source of the ultrafast HHG suppression is the enhanced generation of carriers due to two-photon excitation, which becomes strongly enhanced when the two pulses are present on the sample in the same time. The higher excited carrier density then leads to stronger suppression of the HHG via decoherence due to electron-electron scattering. To resolve the underlying physical mechanism we study the dynamics of the HHG yield for different detuning of the near-resonant excitation from the exciton energy (see Fig. \ref{fig:hhgtime}b-d). Fig. \ref{fig:absorption}a shows the relative suppression of the harmonic yield  as a function of pump peak intensity for three different near-resonant photon energies of the excitation pulse. The suppression is stronger as the detuning becomes larger. Therefore, the HHG suppression cannot be explained by the OSE, for which the energy shift is indirectly proportional to the detuning energy. Another argument is based on the fact that the Stark energy shift is of the order of only several tens of meV\cite{slobodeniuk2022semiconductor}. The estimated decrease of the tunnelling probability induced by such small increase of the band gap is only few percent, which is much less than the observed relative change of the HHG yield.

The observed ultrafast suppression of the HHG yield with near-resonant pump is thus caused by two-photon absorption and subsequent electron dephasing. The excitation process is illustrated in Fig. \ref{fig:absorption}b together with the two-photon absorption coefficient in MoS\textsubscript{2} monolayer calculated by the theory presented in the following chapter. The excitonic levels \cite{splendiani2010emerging,niu2018thickness,frisenda2017micro} form clear peaks in the two-photon absorption spectra. For the near-resonant pulse at 1.46 eV, the sum of photon energies for two-photon transition (infrared HHG driving pulse at 0.6 eV) gives 2.06 eV, which is partially overlapped with 1sB exciton resonance at E\textsubscript{B}=2.03 eV. For the higher photon energies of  1.55 eV and 1.65 eV, the sum of the photon energies is shifting away from 1sB exciton resonance to the region, where the two-photon absorption coefficient is decreasing. This explains, why the measured HHG suppression is stronger for lower pump photon energies corresponding to larger detuning from the exciton resonance. 

Recently, similar ultrafast suppression of HHG yield has been measured in bulk ZnO\cite{xu2022ultrafast}. The phenomenon was interpreted by a decrease of the intensity of the infrared pulse, which is partially absorbed in the crystal due to frequency mixing processes. The lower intensity of the infrared pulse should then lead to a suppression of the HHG. In our experiments with MoS\textsubscript{2} monolayer, any propagation effects can be excluded. The origin of HHG suppression in the time overlap region is the creation of electron-hole pairs by two-photon absorption of the weak near-resonant and strong off-resonant infrared pulses. This effect is not present in ref. \cite{xu2022ultrafast}, where the sum of photon energies of the two pulses is not sufficient to create electron-hole pairs in ZnO via two-photon absorption and only multiphoton processes are available. 


\section{Theory of two-photon absorption in MoS\textsubscript{2} monolayer}

We consider 7-band model of the TMD monolayer which goes beyond the simplest two-band case (see Supporting Information for details). The rate of optical transitions, induced simultaneously by two photons can be written as
\begin{align}
\Gamma(f,in)=\frac{2\pi}{\hbar} |\langle f|U|in\rangle|^2\delta(E_f-E_{in}-\hbar\omega_1-\hbar\omega_2).
\end{align}
Here $\langle f|U|in\rangle$ is the matrix element of two-photon process. It couples the initial 
state  $|in\rangle=|0\rangle$ (with occupied valence bands and empty conduction bands)
and final exciton state $|f\rangle=|\mathbf{q},n,\tau,c,v\rangle$ (with the total momentum 
$\mathbf{q}$, discrete quantum number $n$) consisting of an electron in conduction ($c$) band 
and a hole in valence ($v$) band of $\tau=\pm 1$ valley. We skip the bands' spin indices here, 
because optical transitions conserve spin, i.e., couples only the bands with the same spin. 
Here, the operator $U$ is responsible for the two-photon process.
It corresponds to the second order processes in perturbation theory, see \cite{Landau, Sakurai}. 
The perturbation operator is taken in the form $H_\text{int}(t)=-\mathbf{P\cdot E}$, 
where $\mathbf{P}$ is the monolayer's polarization operator and $\mathbf{E}$ is the in-plane electric field
of incoming light pulses. Taking into account that optical pulses in the experiment are linearly polarized with
polarization vectors $\mathbf{e}_1$, $\mathbf{e}_2$, field amplitudes $|E_1|$, $|E_2|$, and phases 
$\phi_1$, $\phi_2$ we write
\begin{equation}
\begin{split}
\mathbf{E}=&\mathbf{E}_1\cos(\omega_1t+\phi_1)+\mathbf{E}_2\cos(\omega_2t+\phi_2)= \\
=&(|E_1|e^{i\phi_1}e^{i\omega_1t}+|E_1|e^{-i\phi_1}e^{-i\omega_1t})\mathbf{e}_1
\\+&(|E_2|e^{i\phi_2}e^{i\omega_2t}+|E_2|e^{-i\phi_2}e^{-i\omega_2t})\mathbf{e}_2 \\
=&(E_1^*e^{i\omega_1t}+E_1e^{-i\omega_1t})\mathbf{e}_1+
(E_2^*e^{i\omega_2t}+E_2e^{-i\omega_2t})\mathbf{e}_2.
\end{split} 
\end{equation}
For this particular case the interaction term takes the form 
\begin{equation}
\begin{split}
H_\text{int}(t)=-\mathbf{P\cdot e}_1(E_1^*e^{i\omega_1t}+E_1e^{-i\omega_1t})
\\-\mathbf{P\cdot e}_2(E_2^*e^{i\omega_2t}+E_2e^{-i\omega_2t}). 
\end{split}
\end{equation}
Using the results of Refs.~\cite{Landau, Sakurai} we obtain the following expression for the transition rate 
$\Gamma_{\mathbf{q}',m}$ to the excitonic state with energy $E_m(\mathbf{q}')$ due to two-photon absorption 
\begin{align}
\label{eq:absorption_rate}
\Gamma_{m,\mathbf{q}'}=\frac{2\pi}{\hbar}|E_1|^2|E_2|^2\Big[\sum_\tau|M_{\mathbf{q}',m,\tau}|^2\Big]
\delta\big(E_m(\mathbf{q}')-\hbar[\omega_1+\omega_2]\big),
\end{align}
with the two-photon absorption amplitude 
\begin{equation}
\begin{split}
\label{eq:absorbtion_amplitude}
M_{\mathbf{q}',m,\tau}=\sum_{\nu}\Big[&\frac{\langle\mathbf{q}',m,\tau,c,v|(\mathbf{P^\tau\cdot e}_1)|\nu\rangle\langle\nu|(\mathbf{P^\tau\cdot e}_2)|0\rangle}{E_\nu-\hbar\omega_2}\\+&\frac{\langle\mathbf{q}',m,\tau,c,v|(\mathbf{P^\tau\cdot e}_2)|\nu\rangle\langle\nu|(\mathbf{P^\tau\cdot e}_1)|0\rangle}{E_\nu-\hbar\omega_1}\Big].
\end{split}
\end{equation}
Here, the index $\nu$ denote all the information about the excitonic states, i.e. $\nu=\mathbf{q},n,\tau,c+j,v-l$. 
Here $E_\nu=E_{n}(\mathbf{q}')$ is the energy of the exciton in $\tau$ valley with momentum $\mathbf{q}$, quantum number $n$, consisting of an electron in $c+j$ ($j=0,1,\dots$) conduction and a hole in $v-l$ ($l=0,1,\dots $) 
valence bands, respectively. 

Using this approach we derive the expression for the two-photon absorption coefficient (details can be found in Supporting Information). 

\begin{align}
&\beta(\omega_1,\omega_2)\approx 256\pi^3\alpha^2\frac{(\omega_1+\omega_2)^3}{\omega_1^2\omega_2^2} \nonumber \\
&\times\left[\frac{\gamma_6\gamma_4^*A}{(E_{c+2}-E_v)(E_{c+2}-E_c)}
+
\frac{\gamma_5\gamma_2^*B}{(E_c-E_{v-3})(E_v-E_{v-3})}\right]^2 \nonumber \\ &\times
\sum_{m=1}^\infty \frac{1}{(m+\delta)^4}\delta(E_m(0)-\hbar\omega_1-\hbar\omega_2).
\end{align}

Here $\alpha=e^2/\hbar c$ is the fine structure constant. The parameter $\beta(\omega_1,\omega_2)$ can be associated 
with the absorption coefficient for the monolayer TMD. To evaluate the numerical value of this coefficient 
one needs to know: the $\mathbf{k\cdot p}$ coupling parameters $\gamma_2,\gamma_4,\gamma_5,\gamma_6$, the 
energies $E_{v-3},E_v,E_c, E_{c+2}$ (see Ref.~\cite{Durnev2017}); 
numerical values of $A$, $B$, and $\delta$ parameters, which can be evaluated numerically using $\mathbf{k\cdot p}$ 
parameters \cite{Durnev2017} and the two-body approximation (see Supplementary data in \cite{Kipczak2023}); 
and the spectrum of excitons $E_m(0)$ in monolayer TMD \cite{Molas2019}. 

To estimate the absorption coefficient we can use the semi-analytical formula for the spectrum of the excitons 
\cite{Molas2019}  
\begin{align}
E_m(0)=E_g-Ry^*\frac{\gamma}{(m+\delta)^2}, 
\end{align} 
where $E_g$ is the single-particle band gap in the system, $Ry^*=\mu e^4/(2\hbar^2\varepsilon^2)$ is the effective Rydberg energy, for the exciton with reduced mass $\mu$ and dielectric constant $\varepsilon$ of the medium surrounded the monolayer. Dimensionless parameters $\gamma, \delta$ depend on the ratio of the effective Bohr radius 
$a_0^*=\hbar/m_0e^2$ and effective in-plane screening length of the monolayer $r_0^*=r_0/\varepsilon$. 
Taking into account the parameters for the considered system $\varepsilon=1.6$, $\mu=0.26m_0$, $r_0=41.5\mbox{\AA}$, 
the energy positions of the $1s$ states for $A$ and $B$ excitons $E^A_{1s}=1.886$~meV and $E^B_{1s}=2.032$~meV, respectively \cite{slobodeniuk2022semiconductor} we obtain $Ry^*\approx 1.38$~eV, $\gamma\approx 0.943$, 
$\delta\approx 0.746$. Using these parameters we calculate the spectrum of the excitons. 
Then using the Elliott type formula (6) we estimate the absorption coefficient for the first 5 $s$ 
excitonic states (as an example) and broadened delta function with $\Gamma=0.026$~meV 
and present it in the Fig. \ref{fig:absorption}b.

\section{Conclusion}
In conclusion, we experimentally investigate the ultrafast modulation of high harmonic generation in 2D transition metal dichalcogenide MoS\textsubscript{2} using resonant and near-resonant light. We show that the pre-excitation of high density carriers leads to ultrafast dephasing and suppression of coherent interband polarization, which is responsible for macroscopic HHG. With the resonant pre-excitation, the dynamics of HHG suppression follows the recombination of excitons in the material. With the near-resonant pump pulse with photon energy lower than the exciton transition energy, the excitons are generated mainly via two-photon absorption driven by a combination of one photon from the pump pulse and one photon from the strong infrared driving pulse. This process occurs only when the pulses are overlapped in time on the sample. Combined with the short duration of pulses used in this study, this leads to ultrafast modulation of HHG at sub-100 fs timescales, thus enabling a new class of fast nonlinear optical devices working in the strong-field regime.

\section*{Supplementary material}
See the supplementary material for more details on the theory of two-photon absorption in MoS\textsubscript{2} monolayer.

\begin{acknowledgments}
The authors would like to acknowledge the
support by Charles University (UNCE/SCI/010, SVV2020-260590, PRIMUS/19/SCI/05, GA UK 1190120) and Czech Science Foundation (23-06369S). Co-funded by the European Union (ERC, eWaveShaper, 101039339). Views and opinions expressed are however those of the author(s) only and do not necessarily reflect those of the European Union or the  European Research Council. Neither the European Union nor the granting authority can be held responsible for them. 
\end{acknowledgments}

\section*{Author Declarations}
\subsection*{Conflict of Interest}
The authors have no conflicts to disclose.
\subsection*{Author Contributions}
\textbf{Pavel Peterka:} Investigation (equal), Formal analysis (lead), Writing – original draft (lead)., \textbf{Artur O. Slobodeniuk:} Methodology (equal), Software (equal), \textbf{Tomáš Novotný:} Methodology (equal), \textbf{Pawan Suthar:} Methodology (equal), \textbf{Miroslav Bartoš:} Resources (lead), \textbf{František Trojánek:} Software (equal), \textbf{Petr Malý:} Writing – review \& editing (supporting), \textbf{Martin Kozák:} Conceptualization (lead), Investigation (equal), Methodology (equal), Writing – original draft (supporting), Writing – review \& editing (lead), Supervision (lead).

\section*{Data Availability Statement}

The data that support the findings of
this study are available from the
corresponding author upon reasonable
request.

\section*{References}
\bibliography{aipsamp}

\end{document}